# Comparing the suitability of Lithium ion, Lithium Sulfur and Lithium air batteries for current and future vehicular applications

Gorazd Lampič,[1]* Gorazd Gotovac,[1] Hugh Geaney,[2,3] Colm O'Dwyer[2,3]*

[1] *Elaphe Propulsion Technologies Ltd., Tesova 30, 1000, Ljubljana, Slovenia*
[2] *Department of Chemistry, University College Cork, Cork T12 YN60, Ireland*
[3] *Micro-Nano Systems Centre, Tyndall National Institute, Lee Malting, Cork T12 R5CP, Ireland*
* Correspondence: G. Lampič, email: gorazd@elaphe-ev.com; C. O'Dwyer, email: c.odwyer@ucc.ie; Tel: +353 (0)21 4902732, Fax: +353 (0)21 4204097

**Abstract:** In this report, future performance demands of batteries for various vehicular applications are modelled. Vehicles ranging in size from electric bikes to heavy trucks are assessed using driving cycle data which allows key performance parameters such as desired range (km), specific energy of the battery (Wh/Kg), cycle life requirement and expected price per unit capacity (Euro/kWh) to be calculated. These projected performance requirements are compared with the outputs for three existing Li-ion batteries (namely (a) Kokam based high specific energy source (b) A123 based high power energy source and (c) Winston low cost system). The theoretical, current state of the art and projected performance parameters for 'beyond Li-ion' technologies (Li-S and Li-O$_2$) are also compared to the modelled battery performance demands. The analysis indicates that current battery technologies are unlikely to meet future requirements in terms of required specific energies and will likely be too costly. In comparison, fully realized beyond Li-ion alternatives may deliver the required specific energy for the full range of vehicles examined. However, scale-up of these systems is a daunting challenge and their successful implementation will depend on improvements in terms of cycle life, electrode and electrolyte stability, rate performance and development of practical battery architectures.

**Keywords:** Battery; electric vehicle; Li-O$_2$ battery, Li-S battery, Li-ion battery.

## 1. Introduction

The electrification of worldwide vehicular fleets has been identified as a key means of reducing global reliance on exhaustible fossil fuels.[1-6] To date, great strides have been made in the development of vehicles where the internal combustion (IC) engine is either entirely replaced by batteries in fully electric vehicles (EVs) (e.g. Nissan Leaf) or by hybrid systems (e.g. Toyota Prius) where batteries are used in conjunction with smaller IC engines.[7,8] However, EV performance depends heavily on the energy source which is the main bottleneck for range, weight and price (which in turn are the three main performance indicators). The ultimate aim is to create cheaper battery systems with higher specific energies which can significantly extend the mileage range per charge and remove the hurdle of range anxiety currently associated with fully EVs.[1,9-11] Ultimately, plug-in EVs having comparable mileages to conventional IC vehicles are desired but significant improvements are required in terms of battery output. Towards this aim, a wide range of battery technologies have been investigated from fundamental material science level to full battery module scale tests. These batteries have taken the form of nickel metal hydride (NiMH),[12] a wide range of Li-ion chemistries[9,13-17] and so called 'beyond Li-ion' (i.e. Li-S,[18] Li-O$_2$[19-23] and other metal-O$_2$ batteries[24,25]) systems. As a result of these ongoing efforts, several new battery technologies in Technology Readiness Level (TRL) 7 and above may be expected in the future.

Despite the promise shown by these emerging technologies, the issues associated with upscaling promising lab based results to full vehicle level cannot be overlooked. Lab scale results are typically quoted in terms of mAh g$^{-1}$ where the weight considered for the calculation is often the 'active' material. Very often, conductive additives, binders and current collectors (even though they may participate in electrochemical reactions in certain cases [26,27]) are ignored when determining the reported gravimetric capacities. This can make it difficult to assess the practical potential of different electrode materials. This issue may also be exacerbated by the use of low mass loadings which typically leads to very high gravimetric capacities but low areal capacities. Very often, operation of battery electrodes with increased mass loadings does not lead to a proportional increase in the capacity noted due to mechanical issues such as delamination of the active material.[28]

'Beyond Li-ion' technologies may be more problematic in real world applications than more established Li-ion systems due to their reliance on metallic Li (or even more reactive alkali metals e.g. Na[25] or K[29]) as anodes materials, increased architectural complexity and typically poorer cycle life. In the case of Li-O$_2$ batteries for example, the often quoted theoretical specific energy of 3458 W h kg$^{-1}$ is unlikely to be achieved in practical systems. One of the primary hurdles to this technology is the need to provide pure $O_2$ to the cathode (due to side reactions caused by the presence of $CO_2$[30,31] and/or $H_2O$[32,33] in air) which incurs a major weight penalty. To date, parasitic by-products have been seen as ubiquitous in the Li-$O_2$ system. These complex side reactions due to a range of issues including cathode instabilities,[34] catalyst driven side reactions,[32,35,36] electrolyte instabilities[37-39] and perhaps most importantly, the reactivity of $Li_2O_2$ and its intermediates.[19] In an effort to limit the impact of these issues, the majority of recent lab-scale Li-$O_2$ reports have assessed cycle life at limited depths of discharge (typically at 1000 mAh/g for the active material) using pure $O_2$ and impractically low mass loadings for the cathode (often of the order of 1 mg/cm$^2$ or lower) which make it difficult to gauge their prospects in practical applications. Li-S batteries seem closer to industrial readiness in terms of cycle life but exhibit issues such as complex discharge chemistry (involving various polysulphide species and the possibility of associated shuttling between the cathode and anode), and the aforementioned reliance on a Li anode, with real world Li-S batteries likely to operate at below theoretical capacities.[18,40-43]

The purpose of this report is to analyse the requirements and characteristics of batteries suitable for EVs and compare their current performance with the existing EV market in different vehicle types and segments. Additionally, the analysis offers predictions and requirements for existing and expected future battery development to suit a range of EVs under different driving conditions. The key parameters and indicators describing the battery characteristics are specific energy, price per energy unit, continuous and peak C-rate, and cycle life. Commercial and emerging battery types (Li-S, Li-$O_2$ and best-in-class Li-ion systems) and parameters of a range of vehicles (weight, number of wheels, wheel diameter etc.) are considered within the time period from 2010 until 2050. Ten different vehicle types (ranging from a scooter to heavy truck) were analysed from an energy and power demand perspective and the required battery characteristics were detailed. Existing battery characteristics (of commercially available cells) are presented and compared with those envisioned for future vehicles. Based on this, the expected market opportunities for vehicles and for electric vehicles is described.

## 2. Models and Methods

The analysis includes ten different vehicle types, based on projected vehicle parameters from 2010 until 2050. For each vehicle type and point in time, vehicle parameters are selected and battery characteristics calculated. The following vehicles are presented: e-bike, e-scooter, urban car, compact car, high end car, sports car, city bus, delivery van, middle size truck and heavy truck. The approximate vehicle characteristics are taken for each of the vehicle types. It has to be pointed out that there are relatively big differences even within each of the vehicle types but these data are used as illustrations. The most important vehicle parameters included in the simulations are shown in Table 1 with example calculations for three different cars therein. The simulations allow the different battery requirements to be calculated. Among the most important calculated battery parameters are:

- Specific Energy [Wh/kg]
- Continuous C-rate
- Peak C-rate
- Price Per Capacity [EUR/kWh]
- Cycle Life
- Range [km]

**Table 1.** Example calculations showing the vehicle parameters for "Urban car" in year 2020, electric bike in year 2050 and compact car in year 2030.

| Parameter | Urban car in 2020 | e-bike in 2050 | Compact car in 2030 |
|---|---|---|---|
| Full vehicle mass [kg] | 675 | 110 | 1100 |
| Frontal area surface [m$^2$] | 2 | 1.2 | 2.4 |
| Coefficient of air drag | 0.275 | 0.6 | 0.26 |
| Coefficient of rolling resistance | 0.02375 | 0.02 | 0.0225 |
| Wheel radius [m] | 0.25 | 0.33 | 0.25 |
| Number of drive motors | 2 | 1 | 2 |
| Regenerative braking efficiency | 0.55 | 0.7 | 0.6 |
| Energy production and charging efficiency | 0.825 | 0.9 | 0.85 |
| Motor and controller efficiency | 0.825 | 0.9 | 0.85 |
| Vehicle mass without propulsion | 275 | 10 | 700 |
| Hotel load [W] | 1225 | 50 | 1400 |
| Max speed [km/h] | 105 | 35 | 140 |
| Continuous speed [km/h] | 77.5 | 30 | 120 |
| Max hill climbing ability [%] | 22.5 | 20 | 25 |
| Middle grade hill [%] | 10 | 10 | 10 |
| Speed in max grade hill [km/h] | 40 | 10 | 30 |
| Speed in middle grade hill [km/h] | 75 | 15 | 60 |
| Range [km] | 225 | 60 | 350 |
| Zero emission range [km] | 225 | 60 | 350 |
| Speed for acceleration A [km/h] | 80 | 20 | 80 |
| Speed for acceleration B [km/h] | 50 | 10 | 50 |
| Acceleration time from 0 to A [s] | 8.5 | 3 | 9 |
| Acceleration time from 0 to B [s] | 4.75 | 2 | 5.5 |
| Vehicle durability [km] | 225000 | 50000 | 300000 |
| Energy price [EUR/kWh] | 0.15 | 0.3 | 0.2 |
| Hydrogen price [EUR/kg] | 1.5 | 3 | 2 |
| Gasoline price [EUR/l] | 3.625 | 10 | 5.5 |
| Simulation image size | 300 | 300 | 300 |
| Time step in simulation [s] | 0.5 | 0.5 | 0.5 |
| Battery voltage [V] | 225 | 60 | 450 |
| Battery mass [kg] | 175 | 3 | 250 |
| Battery price [EUR] | 2375 | 50 | 3500 |
| Driving cycles (Shown in Fig. 1 a-c) | Fig. 1a | Fig. 1b | Fig. 1c |

The selection of parameter values for each different vehicle is based on a projected change in overall weight, forecast based on weights reductions to the present day. As an example, consider the 'compact car' vehicle type defined as a modelling parameter in this analysis. The full vehicle mass in 2010 is set to 1400 kg, a typical value for a family sedan. In 2050, we analyse battery requirements based on a mass reduction estimate of ~25% to 1100 kg due to the use of lightweight materials, advanced propulsion architectures and weight reduction in all components. The air drag coefficient is also modelled to be reduced from $C_d = 0.3$ to 0.2 as the car profile is modified to improve efficiency [44]. The rolling resistance coefficient is also expected to decrease by 20% due to expected improvements in tyre technology. The driving range is thus expected to increase from 200 to 500 km facilitated by improvements in battery technology – the present work compares the projected improvements in Li-ion and two next-generation technologies regarding range and overall performance for a range of vehicles into the near future. The model parameters (such as weight for example) can be chosen as required. The model used in this analysis is available from the authors.

In order to calculate these parameters several supplementary parameters are also included such as:
- Battery Weight [kg]
- Battery Capacity [Wh]
- Number of Cells
- Cell Capacity [Ah]
- Energy Per km [Wh/km]
- Continuous power [W]
- Peak power [W]

Some of these parameters (range and battery weight for example) are predefined as vehicle characteristics and others are calculated from the remaining vehicle data and expected performance.

## 3. Results and Discussion

Examples of the input parameters are provided in Figs 1 and 2, and Table 1. The former shows driving cycles for three different vehicles used in the calculations while the latter details the characteristics and coefficients used for each of these three vehicles.[45] All of the vehicle characteristics were modelled between 2010 and 2050 in 10-year intervals. For clarity of presentation of example input parameters and results, three arbitrary combinations of vehicle types and time are presented in Table 1. Similar calculations were performed for each of the vehicle types for each time point into the future. From these data the more detailed battery parameters are calculated accordingly. Simulation results for the parameters associated with a 'delivery van' for the year 2040 are presented in Tables 2-4.

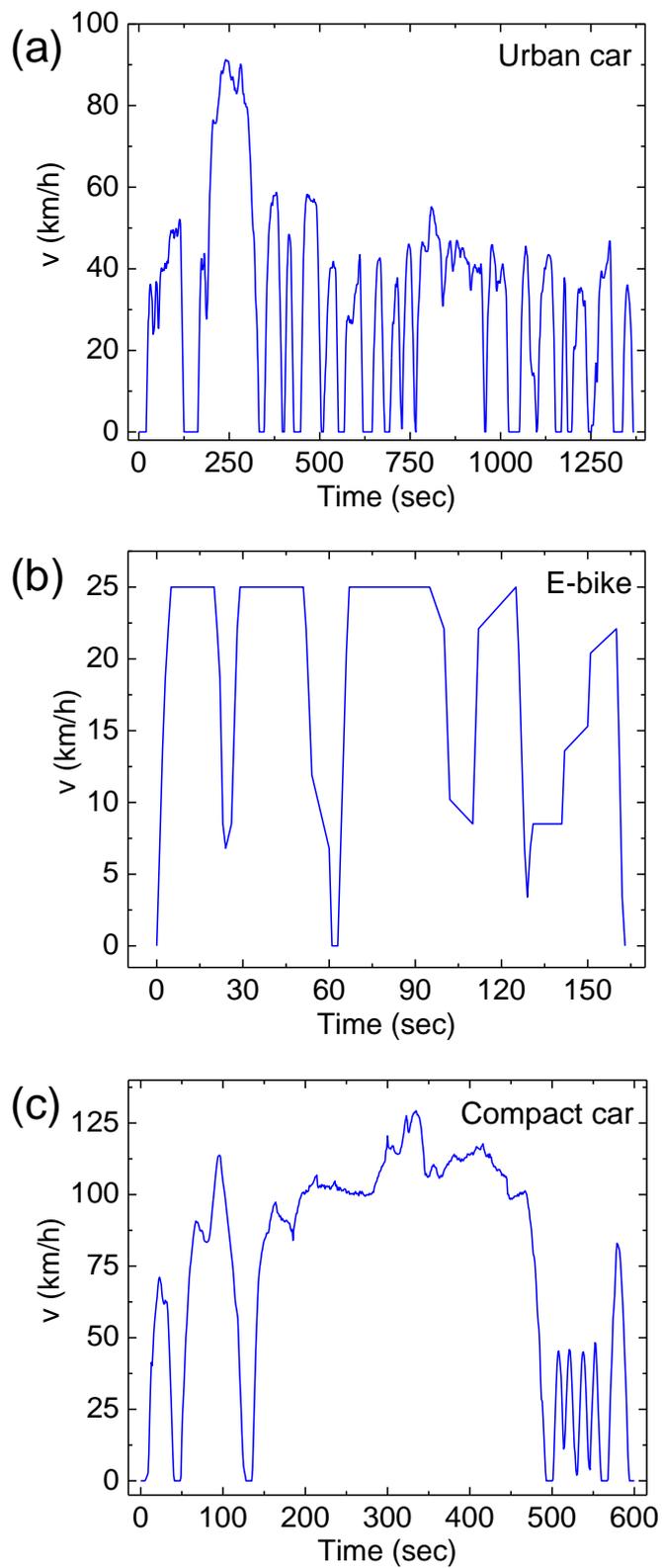

**Fig. 1** Speed vs time profiles used in sample calculations for the following vehicles. (a) Complex driving cycle used for the urban car simulation in 2020. (b) Driving profile up to 25 km/h driving cycle which is used for the e-bike analysis in 2050. (c) Speed vs. time in complex urban driving cycle used for the compact car in 2030.

**Table 2.** Delivery van 2040 driving cycle results.

| Basic cycle data | Value |
|---|---|
| Max torque in cycle [Nm] | 1055 |
| Max torque in cycle (per motor) [Nm] | 528 |
| Torque at top speed (145 km/h) [Nm] | 506 |
| Max power in cycle [W] | 46978 |
| Average drive power in cycle [W] | 6810 |
| Average braking power [W] | -1691 |

**Table 3.** Torque and power calculations with all motors together for delivery van 2040 special driving conditions results.

| Conditions | Grade [%] | Speed [km/h] | Frequency [rev/s] | Torque [Nm] | Power [kW] |
|---|---|---|---|---|---|
| **Steepest grade** | 27.5 | 30 | 4.348 | 1637 | 44.7 |
| **Middle grade** | 10 | 60 | 8.697 | 673 | 36.77 |
| **Top speed** | 0 | 145 | 21.018 | 506 | 66.78 |
| **Continuous speed** | 0 | 122.5 | 17.756 | 397 | 44.33 |
| **Acceleration** | 0 | 0-80 | 0 - 11.60 | 1881 | 68.46 |

**Table 4.** Modelled parameter values for the 'Delivery Van' using projected vehicle parameters in 2040.

| Parameter | Value |
|---|---|
| Range[km] | 412.5 |
| Battery Weight[kg] | 225 |
| Battery Capacity[Wh] | 76553 |
| Number Of Cells | 146 |
| Cell Capacity[Ah] | 142 |
| Energy Per km[Wh] | 185.6 |
| Power cont.[W] | 44717 |
| Power peak[W] | 68460 |
| Specific Energy[Wh/kg] | 340.2 |
| C-rate Cont. | 0.584 |
| C-rate Peak | 0.894 |
| Price Per Capacity [€/kWh] | 117.6 |
| Cycle Life | 1091 |

*3.1 Comparison of Li-ion, Li-S and Li-air battery requirements for vehicles*

The most important parameters of the required batteries are compared for different vehicles and different time points during the time period from 2010 to 2050 in Fig. 2. These values are to be compared with the expected characteristics of future technologies (advanced Li-ion, Li-S and Li-O$_2$) as the technology evolves over time. An overlap of these expectations with the emerging battery performance would mean potential market opportunity for the chemistry. In Fig. 2a the expected ranges for the various vehicle types are outlined. As may be expected, the ranges (km) for all of the EVs are expected to increase over time with a particularly marked range increase required for the heavy truck and high-end car. Comparing Figs 2a and b, we note that the predicted range increase closely follows the projected need for greater battery specific energy for each vehicle type, within associated driving cycles and with a reduction in overall battery pack weight (Fig. 2b).

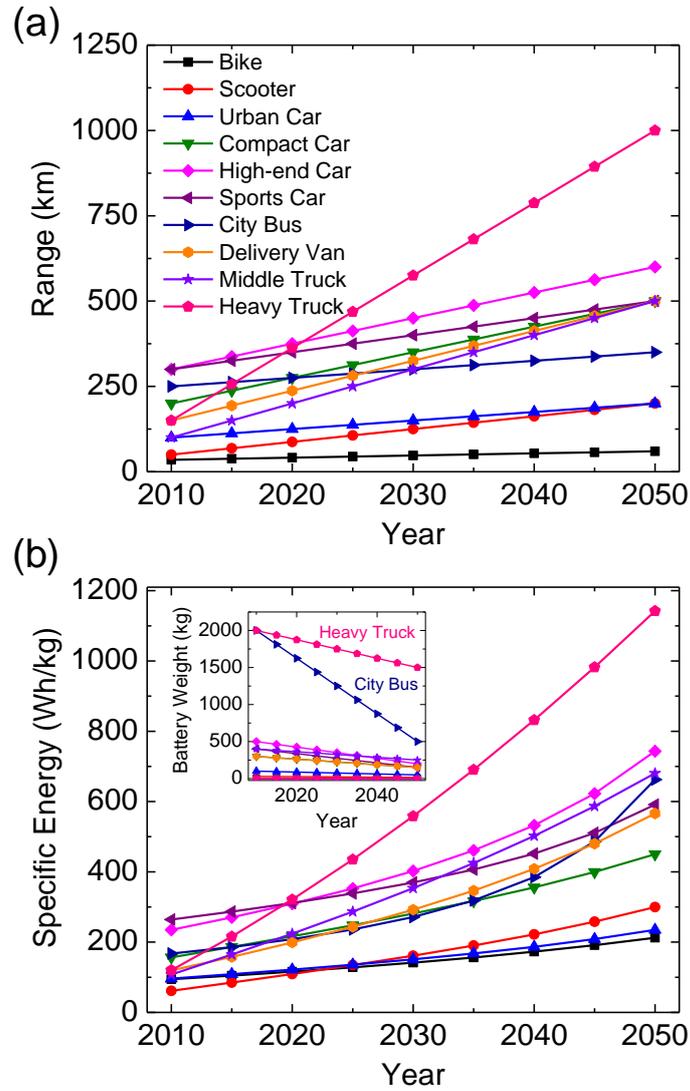

**Fig. 2.** Model output predictions for electric vehicle (a) range and (b) necessary battery specific energies over the projected time period 2010-2050 for all 10 vehicle classes.

While this places demands on the battery chemistry involved (with a near ten-fold increase for the heavy truck and more modest increases of ≤ 3× the 2010 value for specific energy for the other vehicles), it also means that the required cycle life can be reduced. Figure 3a shows the predicted reduction in cycle life demand for all 10 vehicle classes from the models in line with improvements in specific energy. The model predicts that advances in specific energy that provide range increases to facilitate all EV version of major vehicles classes may be advantageous, particularly when taking into account the low cycle life of current beyond-Li-ion batteries.[19,46] The model of vehicle energy and power requirements is robust, since parameters of different fields, such as vehicle data, battery data, development roadmaps, marketing expectations etc. are combined and avoid excessive variables that would make it unclear and less useful. For details, such as voltage drop due to temperature variations or different loads, the model assumes an approximately similar behavior is present in all cases. We note however, that voltage drop, voltage stability vs. SOC, charging time associated with specific energy increases etc. must also been considered in practical EV battery packs in the future, and as discussed below, the C-rate may influence capacity and specific energy. The high cycle lives projected for the heavy truck, middle truck and city bus reflect their high battery lifetime figures (in km). In Fig. 3b, the projected price per unit capacity (€/kWh) is presented as a function of time. It can be seen that a dramatic decrease in the cost per unit capacity is required for all of the vehicles with an expected cost of < 200 €/kWh in all cases in 2050. A marked drop in price (per kWh) has been seen for batteries for EV's in the past decade and this improvement will need to continue to satisfy the requirements outlined here.[47] The simulations above have outlined the battery requirements for current and future vehicles. These values reflect what vehicle manufacturers would require to fulfil their customer needs and expectations. The code/model/data is available from the authors if anyone wishes to do their own calculations. Many of these parameters and equations used in the model are defined in EV textbooks or can be derived bottom up in any case.

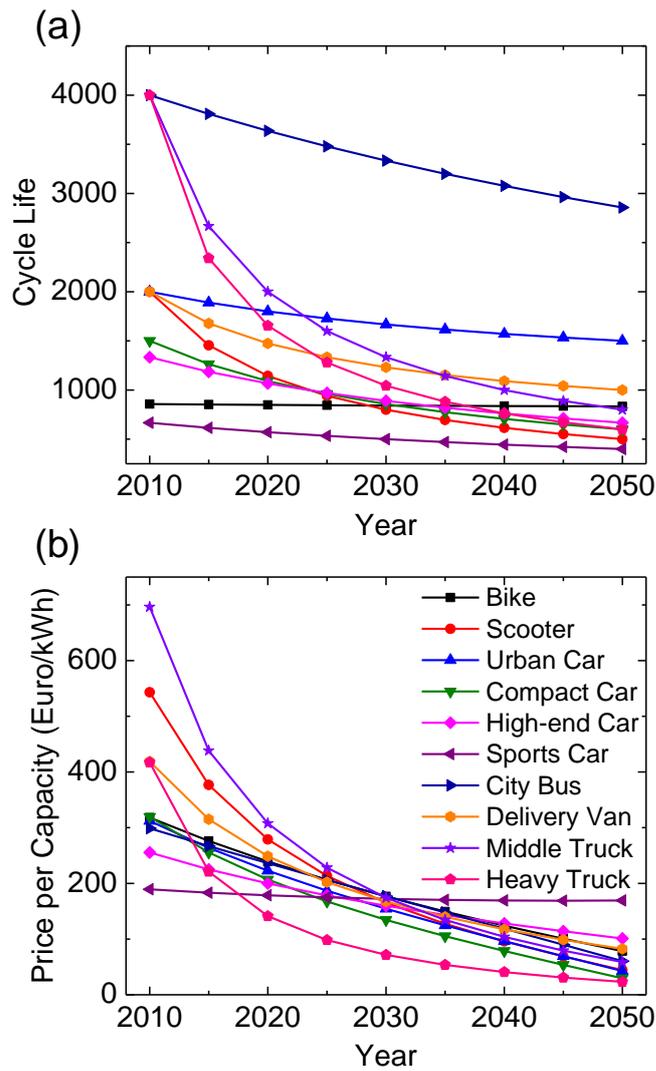

**Fig. 3.** Model output predictions for electric vehicle battery (a) cycle life and (b) price per capacity over the projected time period 2010-2050 for all 10 vehicle classes.

*3.2 Comparison with Existing Products based on Li-ion Technology*

Among the many existing products, three common batteries have been analysed here.
- Kokam lithium polymer cells
- Winston LiFePO$_4$ cells
- A123 LiFePO$_4$ cells

Among these batteries there are different classes in respect to high power or normal power application, however, to keep the analysis as simple as possible, the data of the most commonly used cells of those manufacturers is used. The parameters are compared at cell level (Table 5) and at complete energy source system level (Table 6). The latter includes the cells, battery box, BMS, fuses and all connections between those elements. The charger and possible cooling systems are not part of the analysis.

**Table 5.** Existing cell level data for current commercial Li-poly and Li-ion cells.

| Parameter | Kokam cell – high energy (SLPB70460330) | Kokam cell – high power (SLPB0460330H) | Winston cell (GWL/Power SP-LFP100AHA) | A123 cell |
|---|---|---|---|---|
| Specific energy [Wh/kg] | 161 | 137 | 103 | 104 |
| Continuous C-rate | 1 | 5 | 1 | 28 |
| Peak C-rate | 3 | 8 | 5 | 48 |
| Price [€/kWh] [48] | 600 | 600 | 250 | 450 |
| Cycle life | 800 | 800 | 2000 | 1000 |

**Table 6.** Existing Li-ion battery system data and comparison with requirements from performance modelling.

| Parameter | Kokam based high specific energy source | A123 based high power energy source | Winston low cost system | Range of values for overall 2015 requirements (from Figs. 1 - 3) |
|---|---|---|---|---|
| Specific energy [Wh/kg] | 150 | 80 | 90 | ≈100 – 300 |
| Continuous C-rate | 1 | 28 | 1 | ≈0.5 – 1.7 |
| Peak C-rate | 3 | 48 | 5 | ≈1 – 3.2 |
| Price [€/kWh] | 800 | 650 | 300 | ≈190 – 475 |
| Cycle life | 800 | 1000 | 2000 | ≈600 – 3800 |

The current performances of the three analysed battery systems in Table 6 for 2014/2015 are compared to the driving range values predicted from the modelled vehicles presented in Figs 2 and 3. Here, vehicles are modelled using the range of parameters outlined in Tables 1-4, and applied to model the 10 different classes of vehicle (see Fig. 2), using predictions for vehicle parameters predicted between known values in 2010 and expected variations up to 2050.

From the data it can be seen that the majority of the requirements are just met or are below the desired threshold. Particularly, specific energies of the energy systems are below that envisioned at the upper bound of the requirements. The energy systems are also typically more expensive than the desired prices (note, high performance sports cars based on EV technology are already available and considerably exceed market requirements, thus the price per capacity remains relatively constant over the modelled timespan).

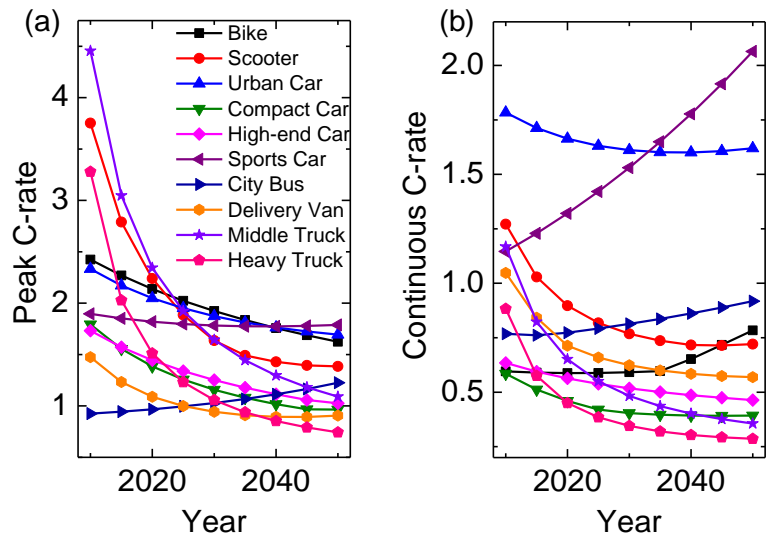

**Fig. 4.** Predicted variation in (a) Peak C-rate and (b) Continuous C-rate requirements over the projected time period 2010-2050 for all 10 vehicle classes.

For the C-rate values, the existing solutions are mostly suitable in the range C/2 to 1 C. While C-rate influence specific energy, the C-rate values predicted in Figs 4a and b are those associated with specific energy values in Fig. 1. These values meet or exceed usable values benchmarked at a sustained <C/2 rate for a >300 mile (480 km) range and a peak C-rate between 1-2 C. Current batteries largely meet the desired cycle lifetime performance aside for the requirement of a city bus, which is much higher than the requirements for other vehicles, however the city bus driving requirements on a daily basis are among the most demanding scenarios for EV systems. The performance indicators for the vehicle types provide some insight into the suitability of different energy systems. Table 6 however, considers the best-case scenario for vehicle requirements with respect to the optimum performance for all parameters of each battery in the analysis, and the representative system do not include higher voltage lithium batteries that provide marginally higher specific energies.

The data summarised in Table 7 demonstrate that e-bikes, scooters and urban cars are projected to be the most likely to be market-ready with performance that approaches design expectations, as predicted by the vehicle parameters in this model. The real market numbers certainly confirm those expectations with over 10 million units annual globally after 2005 and with over 700,000 units in EU in 2010. The next group of vehicles that miss just one of the parameters are the compact car (the Kokam based high specific energy source meets the specific requirement), delivery van, and city bus. The middle size and heavy truck, as modelled, do not have capable energy solutions in three of the five categories based on currently available battery technologies, which explains the gap in the battery market for fully electric large vehicles (some hybrid vehicles do exist).

**Table 7.** Assessment of suitability of existing energy systems to different vehicle classes. The suitability of each energy system (a, b and c) is given for each parameter (x denotes no acceptable system currently available).

| | a) Kokam high specific energy source | b) A123 high power source | c) Winston low cost system | Suitable vehicles (2015 parameters) | | | | | | | | | |
|---|---|---|---|---|---|---|---|---|---|---|---|---|---|
| | | | | Bike | Scooter | Urban | Compact | High-end | Sports | City bus | Delivery | Middle | Heavy |
| Specific energy [Wh/kg] | 150 | 80 | 90 | a, b, c | a, b, c | a, b, c | a | x | x | a | a | a | a |
| Continuous C-rate | 1 | 28 | 1 | a, b, c | | | | | | | | | |
| Peak C-rate | 3 | 48 | 5 | a, b, c | | | | | | | | | |
| Price [€/kWh]* | 800 | 650 | 300 | c | c | c | c | x | x | c | c | c | x |
| Cycle life | 800 | 1000 | 2000 | a, b, c | c | c | c | c | a, b, c | x | c | x | x |

*3.3. Future expectations of emerging beyond-Li ion battery systems*

There are two main future technologies that will be analysed using the vehicle drive cycle and expected performance parameters. The Li-S battery chemistry (outlined in Table 8) is closer to the market while Li-O$_2$ offers higher specific energy but is not as close to commercialization (shown in Table 9).[18,40] The comparison is made in terms of theoretical characteristics, using the state of the art performance indicators in 2014 and speculative expected performances in 2030. It should be noted that the state of the art parameters in 2014 are based on lab-scale experiments of these battery technologies, and as a result, the actual 'active' material mass loadings on electrodes are orders of magnitude from practical cells. As a result, a range of values is provided where possible to present an accurate representation of the current state of the field. For Li-O$_2$ batteries, the majority of cycle life tests are currently being conducted at a limited depth of discharge (1000 mA/g) which can often be a fraction of the total depth of discharge. This testing protocol avoids side reactions due to electrolyte and cathode decomposition associated with complete discharge and recharge.[49] Working at such a limited depth of discharge would likely negate the benefit of the Li-$O_2$ system (high specific capacity) compared to standard Li-ion batteries. Ideally, more stable electrolyte/cathode pairings will be identified in future which allow the realization of full depth of discharge operation. It remains to be seen if a practical Li-$O_2$ with a genuine performance improvement compared to current Li-ion batteries will be developed and they must remain to be seen as a high-risk alternative until practical, scaled up systems can be presented.[50] For both of these beyond Li-ion technologies, theoretical predictions have been made based on the use of a Li metal anode which may not be practical due to performance and safety issues. As a result, alternative anodes which would severely impact the achievable energy density of a full cell due to their increased weights compared to Li have already been proposed.[51,52] The performance of Li-$O_2$ batteries in particular has been shown to be extremely sensitive to temperature (which has a large impact on discharge and charge potentials) and must be considered for practical applications.[53,54]

The future characteristics of the batteries under development and their applicability to EVs can be assessed in Tables 8 (Li-S) and 9 (Li-$O_2$). It is evident that the majority of the theoretical parameters are well above what is required for these applications with specific energy being an obvious example. In the case that these theoretical specific energies could be achieved at a full module basis (i.e. for Li-$O_2$ and Li-S batteries), vehicle designers would have the opportunity to decrease the battery size since the offered range would be satisfactory. This may have repercussion on the improvement or stability of the continuous and peak C-rates at operational DOD, since these values might be relatively low for the new batteries under development.[55] A similar issue is likely to exist with cycle life.[40] If ultra-high capacity batteries (i.e. >750 Wh/kg) can be developed in future, vehicles could be produced with higher ranges than required, meaning that specific power and cycle life could replace range as the most limiting factor for some vehicle types. One possible solution to this problem would be the addition of capacitors or high power cells to mitigate any other performance limitations at peak C-rate operation.[56] Overall, price reduction is the main goal of battery development. In addition, not only cheaper technologies, but also more efficient models of battery and/or vehicle use and ownership have to be developed in order to reduce the complete cost of ownership.

Table 8. Comparison of theoretical parameters, current state of the art, an example commercial Li-S battery and 2030 speculation for Li-S system compared with the range of performance output needs between 2015-2030 shown in Figs 2 and 3.

| Parameter | Li-S Theoretical | Current Li – S state of the art | Current and projected performance (2019 Q1) for commercial Li-S product [23] | Speculative Li – S in 2030 | Range of values for 2015-2030 requirements (from Figs. 2 and 3) |
|---|---|---|---|---|---|
| Specific energy [Wh/kg] | 2,567[18] | 400-500[57] 600[58] | Up to 300 (500) | 800-1500 | ≈100 -575 |
| Continuous C-rate | Not calculated | 0.2[59] | Not given | 1 | ≈0.4 – 1.7 |
| Peak C-rate | Not calculated | Not Given | 3 (5) | 5 | ≈1 – 3.2 |
| Price | < 150 US$/kWh [18] | Not known for scaled battery | Not given | < 200 Euro/kWh | ≈85 – 475 Euro/kWh |
| Cycle life requirements and discussion | ≈250[18] cycles required for 100,000 km driving distance at theoretical specific energy above | Current status: 50-1000 [41,59]  >1000 cycles needed at 600 Wh/kg for 100,000 km[18] | 100 (1500 projected) | >600 cycles for (100,000 km)  Longer cycle life may be achievable if fundamental issues can be overcome | ≈600 - 3800 |

Examining the current state of the art performance, for the Li-S system (Table 8) it can be seen that the specific energies reported meet the requirements for practically all of the energy demands for all vehicle types even up to 2030. Of course, this does not factor in scale-up issues associated with the battery chemistry or losses associated with scale up from a single laboratory cell level to a full scale energy system, but is a promising finding. The theoretical price proposed by Bruce et al. (< 150 US$/kWh) would be acceptable for almost all of the modelled vehicles up until 2030, however, a scaled up price of the most promising systems to date is unknown. The majority of reports on Li-S batteries still present cycle life values that are well below what would be needed for practical cells although the number of cycles is improving in recent literature.[40,41] Of course, the required cycle life will be dictated by both the achievable specific energy and the desired range in future but improvement in this regard seems promising as a better understanding of the fundamental battery chemistry is developed.[41,58] The current and projected parameters for a Li-S battery produced by OXIS Energy are also outlined in the table. Notably, the achieved specific energy figures of up to 300 Wh/kg is already higher than the existing Li-ion alternatives presented above with ambitious specific energies of 500 Wh/kg projected for 2019. The current bottleneck for these commercial batteries is the cycle life (100), which is currently below the threshold for all of the vehicle requirements. However, the projected cycle life of 1500 for 2019 is much closer to the expected demands for all but the most demanding vehicle types (city bus and urban car).

Table 9. Comparison of theoretical parameters, closed $O_2$ system modelled by Gallagher et al., the current state of the art from lab scale test and 2030 speculation for Li-$O_2$ system compared with the range of performance output needs between 2015-2030 shown in Figs. 2 and 3.

| Parameter | Li-$O_2$ (non-aq) Theoretical | Current Li-$O_2$ state of the art | Closed Li-$O_2$ system proposed by Gallagher et al. [60] | Speculative Li-$O_2$ in 2030 | Range of values for 2015-2030 requirements (from Figs. 2 and 3) |
|---|---|---|---|---|---|
| Specific energy [Wh/kg] | 3,505[18] | ≈2000[55] (with practical estimate of 670) | ≈300 | 500-1000 | ≈100 -575 |
| Continuous C-rate | Not known | Not known | 1 | 0.5 | ≈0.4 – 1.7 |
| Peak C-rate | Not known | Not known | Not Given | 1 | ≈1 – 3.2 |
| Price | <150 US$ kWh [18] | Not known for scaled battery | ≈150 USD$ kWh h$^{-1}$ | 200 (will likely depend on need for catalysts) | ≈85 – 475 Euro/kWh |
| Cycle life requirements and discussion | ≈200 [18](100,000 km) | Current status: Limited at full depth of discharge conditions. 100 cycles for carbon free cathode[61]  ≈320 needed at 2000 Wh/kg for 100,000 km  ≈957 at 670 Wh/kg | ≈2140 cycles needed at this specific energy for 100,000 km | >600 cycles needed for 100,000 km at 1000 Wh/kg  Cycle life >300 cycles should be achievable assuming that technical challenges overcome (side reactions, air scrubbing, efficient materials electrochemistry etc.) | ≈600 - 3800 |

For the Li-$O_2$ system, the models predict that the theoretical specific energy is well beyond what would be needed for any of the vehicles modelled in Fig. 2 at any time in the analysis up to 2050, by a factor of ~3 which would accommodate large variability in model overall vehicle weight. The current state of the art for specific energy of a Li-$O_2$ cell is ≈2000 Wh/kg, but with a practical estimate of 670 Wh/kg as reported by Lu et al.[55] However, a thorough full scale analysis of different battery systems conducted by Gallagher et al. predicted a much more conservative estimate of ≈300 Wh/kg for a closed Li-$O_2$ system due primarily to the myriad issues associated with scale up of the battery.[60] Specific energy values of ≈300 Wh/kg would still represent an improvement on the currently available Li-ion cells, but it remains to be seen if the improvement would be significant compared to future improved Li-ion batteries, Li-S, or modern Li-ion technologies with Li metal anodes.

The figure would also fall short of the future energy requirements seen in Fig. 2. The projected price of the Li-$O_2$ system is attractive and would meet most of the requirements but may depend on whether precious metal catalysts are required in any practical cell. The USCAR (U.S. DRIVE - Driving Research and Innovation for Vehicle efficiency and Energy sustainability, US Department of Energy) long-term commercialization goals for EV batteries[62] cites 100 US$ per kWh, which in our analysis is feasible for all vehicle types by 2030 based on battery pack requirements. Perhaps the greatest hurdle to the practical use of Li-$O_2$ system is the very poor rechargeability due to ubiquitous side reactions and many other issue associated with system-level scale up.[37] The poor cycle life is seen from the limited cycle life even for carbon free cathodes and dramatic improvements are required if the Li-$O_2$ battery is to become a primary energy source for electric vehicles, and by comparison to existing technologies, several thousand cycles are required over life spans of 10-15 years. While 2140 cycles are predicted for a total driving range of 100,000 km at a specific energy of ≈300 Wh/kg (Table 9), the USCAR range requirements are set at 240,000 km, but for Li-$O_2$ batteries, cycle life requirements are not defined. In terms of C-rate needs, a Li-$O_2$ battery in a closed system with an order of magnitude less specific energy than the theoretical value, a C/2 sustained discharge rate is benchmarked. There are limited tests available on variable C-rate and asymmetric charge-discharge analyses of these batteries and their adoption into cars with charging power from regenerative braking or other power transfer systems will necessitate faster charge capability. To the best of our knowledge, no commercial Li-$O_2$ batteries have been developed to date in spite of some figures of merit being sufficient for the market, and it remains to be seen if the numerous practical issues can be overcome. As a result, the speculative values for 2030 given in Table 9 may have a larger margin of error than the Li-S values presented in Table 8 for Li-S technologies.

## 5. Conclusions

In this work we have detailed the technical and economic characteristics required for battery systems for different vehicle types in the time period from 2010-2050. The analysis explains why electric two wheelers (e-bikes and scooters) are already on the market in large numbers and suggests which other vehicle types will follow to market. The analysis also indicated the limiting battery parameters for more challenging systems such as heavy goods trucks. Typically, the C-rate and cycle life performance of currently available batteries are suitable for future applications, however, the specific energy and price per energy of the cells needs to be improved. At the same time, it is interesting to note that high-end electric cars exist on the market despite the fact that they do not yet meet the price target for the batteries. We speculate that a limited percentage of customers are willing to pay a premium for the benefits of owning such vehicles.

For future 'beyond Li-ion' technologies to become viable alternatives to existing Li-ion batteries, large strides will need to be made at the level of basic understanding of the systems and battery engineering (in terms of scaling up promising lab-scale results). It remains to be seen if systems based on Li anodes will ever become practical for EV-scale use but advances in Li stabilization routes have shown promising results recently for high capacity Li-ion batteries.[63] The issues associated with the translation of promising lab-scale results to a functioning energy system cannot be underestimated. For both Li-S and Li-$O_2$ batteries, state of the art specific energies (when considering full depth of discharge) at lab-scale dwarf existing Li-ion alternatives. In the case of current Li-$O_2$ testing, the majority of systems are investigated at limited depths of discharge (1000 mAh/g) to limit ubiquitous side-reactions. This exacerbates the difficulty in assessing the potential of this system compared to other technologies, and Li-S batteries seem closer to industrialization. The need to provide a pure $O_2$ reactant to the cathode seems to suggest a major weight penalty associated with the shift to a closed $O_2$ system, where air scrubbing or closed $O_2$ recycling will certainly limit energy densities. In both beyond-Li-ion cases, the issue of limited cycle life seems to be among the most important performance challenge that must be overcome, and many important aspects for electric vehicle requirements that lead to predictions of 10-20 years before market place availability can be found elsewhere [21]. C-rate performance for the system is also critical for vehicle that regenerate power from intermittent sources such as braking, with the proviso that battery performance is weighted with the vehicle type and the typical driving cycle during use, and should be assessed in future studies including materials and new battery cell research.

**Acknowledgments**

This research has received funding from the Seventh Framework Programme FP7/2007-2013 (Project STABLE) under grant agreement no. 314508. We acknowledge useful discussions with OXIS Energy regarding Li-S battery systems.